\documentclass[%
 aip,
 amsmath,amssymb,
 reprint,%
]{revtex4-1}

\usepackage{graphicx}
\usepackage{graphics}
\usepackage{float}
\usepackage{dcolumn}
\usepackage{bm}
\usepackage[utf8]{inputenc}
\usepackage[T1]{fontenc}
\usepackage{mathptmx}
\usepackage{color,soul}
\usepackage{amssymb}
\usepackage{amsmath}
\usepackage{mathptmx}  
\usepackage{longtable,tabularx}
\usepackage{xcolor}
\usepackage{subfloat}
\usepackage{multirow}
\usepackage{rotating}
\usepackage{booktabs}
\usepackage{relsize}
\usepackage{nomencl}
\usepackage{etoolbox}
\usepackage{float}
\usepackage{url}
\usepackage{hyperref}
\usepackage[ruled]{algorithm2e}
\usepackage{nomencl}
\usepackage{xcolor}
\usepackage[normalem]{ulem}

\newcommand{\pder}[2][]{\frac{\partial#1}{\partial#2}}

\makenomenclature
\begin{document}

\preprint{AIP/123-QED}

\title{Combined space-time reduced-order model with 3D deep convolution for extrapolating fluid dynamics}
\affiliation{Department of Mechanical Engineering, University of British Columbia, V6T 1Z4, Canada}
\author{Indu Kant Deo}
\email{indukant@mail.ubc.ca}
\author{Rui Gao}
\email{garrygao@mail.ubc.ca}
\author{Rajeev Jaiman}
\email{rjaiman@mech.ubc.ca}

\date{\today}

\begin{abstract}
There is a critical need for efficient and reliable active flow control strategies to reduce drag and noise in  aerospace and marine engineering applications.
While traditional full-order models based on the Navier-Stokes equations are not feasible, advanced model reduction techniques can be inefficient for active control tasks especially with strong non-linearity and convection-dominated phenomena.
Using convolutional recurrent autoencoder network architectures, deep learning-based reduced-order models have been recently shown to be effective while performing several orders of magnitude faster than full-order simulations. 
However, these models encounter significant challenges outside the training data, limiting their effectiveness for active control and optimization tasks. 
In this study, we aim to improve the extrapolation capability by modifying the network architecture and integrating coupled space-time physics as an implicit bias. 
Reduced-order models via deep learning generally employ decoupling in spatial and temporal dimensions, which can introduce modeling and approximation errors. 
To alleviate these errors, we propose a novel technique for learning coupled spatial-temporal correlation using a 3D convolution network. 
We assess the proposed technique against a standard encoder-propagator-decoder model and demonstrate a superior extrapolation performance.
To demonstrate the effectiveness of the 3D convolution network, we consider a benchmark problem of the flow past a circular cylinder at laminar flow conditions and use the spatio-temporal snapshots from the full-order simulations. Our proposed 3D convolution architecture accurately captures the velocity and pressure fields for varying Reynolds numbers. Compared to the standard encoder-propagator-decoder network, the spatio-temporal-based 3D convolution network improves the prediction range of Reynolds numbers outside of the training data.

\end{abstract}

\maketitle

\section{Introduction}
Partial differential equations govern the evolution of state variables in numerous dynamical systems and physical processes \cite{evans2010partial}.
For example, the dynamics of fluid flow are governed by a set of partial differential equations known as the Navier-Stokes equations.
There has been an interest in solving these equations for reliable active flow control strategies \cite{collis2004issues,donovan1998active,joslin2009fundamentals}.
Active flow control strategies reduce noise emission and aerodynamic drag, lowering the energy consumption of large marine vehicles and aircraft \cite{chung2011effectiveness, jackson2017afterbody}.
To solve these partial differential equations, various numerical discretization techniques such as finite-volume or finite-element are frequently used~\cite{leveque2002finite,johnson2009,hughes2000}.
The solution of Navier-Stokes equations using these methodologies becomes prohibitively expensive and intractable for multi-query analysis, design optimization, and control tasks.
To address these limitations, the full-order Navier-Stokes model is replaced with a lower dimension reduced-order model capable of expressing the physical properties of the problem~\cite{quarteroni2015reduced}.

One of the traditional methods for reduced order modelling is the projection-based technique, which assumes that a low-rank approximation can be expressed as a linear combination of basis functions \cite{quarteroni2014reduced}. 
These basis functions are built from a set of full-order model solutions known as snapshots\cite{schilders2008model}. 
Proper orthogonal decomposition\cite{ravindran2000reduced} is one of these approaches, which generates a linear reduced-order model by decomposing a snapshot matrix into principal components or singular-values and employs Galerkin projection\cite{DBLP:journals/corr/CarlbergBA15} for evolving dynamics in this reduced space. 
However, for many real-world scenarios and non-linear phenomena, linear reduced order models struggle to generalize and produce accurate results \cite{unger2019kolmogorov}.
This limitation of the linear model encourages the use of nonlinear reduced order modeling methods such as kernel PCA \cite{mika1998kernel}, auto-encoders \cite{hinton2006reducing}, and diffeomorphic dimensionality reduction \cite{seeger2009advances}, to effectively reduce the dimensionality and construct non-linear reduced-order models. 
Such projection-based methods are effective in dimensionality reduction, however, they rely on underlying governing differential equations and cannot be applied to spatio-temporal data when the equations are unknown. Machine learning algorithms entail data to learn a specific task, making them suitable to leverage spatio-temporal data and to integrate with measurement data.
 Such data-driven learning models are particularly instrumental during the offline-online strategy for active control by integrating with measurement data. While the data-driven model can be trained to learn a reduced representation from the high-dimensional physical data in the offline stage, the reduced model in the online stage can provide real-time and scalable predictions for control and optimization tasks.

In recent years, machine learning has witnessed a resurgence owing primarily to the enormous success of deep learning models in a wide range of applications~\cite{lecun2015deep}.
One such application is reduced-order modeling, in which a deep learning model is used as a black box technique to approximate a physical system \cite{gupta2022three,deo2022predicting}.
Deep neural networks, one of the most popular deep learning models, have proven to be an effective method for modeling physical systems, e.g. data-driven projection method for the Navier-Stokes equations~\cite{yang2016data} and the reduced-order modeling of fluid forces around bluff bodies \cite{miyanawala2017efficient,miyanawala2018low}.
However, it is widely acknowledged that training such models might require a large amount of data. To reduce the amount of data consumed and make them more interpretable, there is a desire to incorporate existing scientific knowledge into these models \cite{goyal2020inductive}. 
This has resulted in the development of physics-guided neural networks, which incorporate domain-specific physical information into the deep learning architecture.

In order to incorporate scientific knowledge in the neural network framework, many deep learning algorithms utilize physical constraints in the loss function\cite{willard2021integrating}.
Explicit physical constraints in the loss function necessitate numerical solution of a partial differential equation which brings back the computational infeasibility of numerical methods \cite{krishnapriyan2021characterizing,raissi2019physics}.
In recent body of works  \cite{deo2022predicting,mallik2022convolutional,gupta2022three}, deep neural networks have been shown to model complex fluid-structure interaction and wave propagation phenomenon without relying on explicit physical constraints or inductive biases in the loss function.
Convolutional recurrent autoencoder network (CRAN)\cite{bukka2020deep,bukka2021assessment,deo2022learning}, is a deep neural network that can be effective for data-driven model reduction and learning of nonlinear partial differential equations. 
The CRAN-based methodology is an entirely data-driven approach in which both the low-dimensional representation of the state and its time evolution are learned via an implicit inductive bias. The methodology alleviates the problem of knowing the underlying differential equation and the computational burden of numerical methods since neural networks with implicit bias can be efficient for inference.   The CRAN methodology has been successfully applied for fluid-structure interaction \cite{gupta2022hybrid} and underwater radiated noise  \cite{mallik2022convolutional}.

From a theoretical viewpoint, the solutions to an unsteady partial differential equation are functions that simultaneously depend on space and time while satisfying the governing equation and its boundary conditions \cite{logan2008introduction}.
The standard methods for solving time-dependent partial differential equations are to first discretize in space and then solve the resulting ordinary differential equation, such as the method of lines \cite{schiesser2012numerical}, or to first discretize in time and then solve the partial differential equation, such as the Rothe's method \cite{kavcur1986method}.
These numerical discretization methodologies have inspired current deep-learning-based reduced-order models such as CRAN \cite{bukka2021assessment,deo2022learning}.
The CRAN architecture captures spatial correlation via convolution layers, projects high-dimensional data to low-dimensional latent space, and learns dynamics on this latent space using long short-term memory cells.
Decoupling the spatial and temporal dimensions in a deep learning-based reduced order model introduces modeling error.
Because of the intrinsic modeling errors, the encoder-propagator-decoder model produces sub-optimal performance in the extrapolation domain that is outside of the training set. 
To alleviate these errors, 3D convolution networks can be employed to learn the combined space-time dynamics. 
Recently, Pant et al. \cite{pant2021deep}  employed a 3D convolution autoencoder and a 3D U-Net for a reduced-order modeling of fluid flow.
Instead of using a different Reynolds number for the validation case, the authors considered the fluid data for the same Reynolds number for training and testing tasks. Furthermore, the study did not include the information about the generalization and the model's performance on the data outside of the training range.

In this paper, we propose a novel 3D convolutional network to capture the inherent spatio-temporal correlation in the PDE without decoupling spatial and temporal dimensions.
We consider a prototypical problem of the flow past a circular cylinder and utilize the space-time snapshots from the numerical solutions of the incompressible Navier-Stokes equations. 
Our combined space-time 3D convolution method extracts spatial-temporal correlation from data and generalizes better in extrapolation domains outside of the training set.
We particularly focus on the model's performance outside of the training range parameter and the extrapolation capabilities of 3D convolution networks. 
We provide the physical interpretation for the superior extrapolation performance when compared to encoder-propagator-decoder architecture.
We empirically show the superior generalization property of our method by presenting results for a wide range of Reynolds numbers outside of the training range.


The rest of the paper is organized as follows. 
 Section 2 presents some mathematical preliminaries and our proposed  3D convolution neural network.
Section 3 introduces the training strategy for our 3D convolution network architecture.
Section 4 presents a numerical analysis of the 3D convolution network for the flow past a circular cylinder.
Finally, Section 5 concludes with a brief discussion of our findings and some directions for future research.

\section{Methodology}
In this section, we first provide a brief review of data-driven reduced-order representation of a dynamical system. We next introduce our combined space-time reduced-order model with 3D deep convolution.
\subsection{Data-driven reduced system}
We begin with formulating a general data-driven learning of a dynamical system
A generic differential equation for a dynamical system can be presented in an abstract form:
\begin{equation}
\begin{aligned}
\pder{t}{\mathbf{U}(\mathbf{X}, t)}&=\mathcal{F}\left(\mathbf{U}(\mathbf{X}, t)\right)\quad &&\text{on} \quad \Omega \times (0, T),\\
\label{eqn:PDE}
\end{aligned}
\end{equation}
where $\Omega$ denotes the
spatial domain, and $\mathcal{F}$ is a generic nonlinear operator describing the time evolution of the system. 
The solution of this partial differential equation is represented by $\mathbf{U}$: $\Omega \times [0, T]  \rightarrow \mathbb{R}$.
The solution field $\mathbf{U}(\mathbf{X}, t)$ is dependent on both space $\mathbf{X}$ and time $t$, and satisfies the equation.
This equation can be solved using numerical discretization techniques (e.g., finite volume or finite element).
Many discretization techniques begin by discretizing the spatial dimension, resulting in a high-dimensional vector $\mathbf{U}_N$, and then evolve the solution over time.
The set of solutions obtained is known as the solution manifold, which is represented by $\mathbf{S_{U}}$ as:
\begin{align}
\mathbf{S}_{\mathbf{U}}=\left[\mathbf{U}_{N}^{(t_{1})}, \ldots, \mathbf{U}_{N}^{({N_{T}})}\right],
\end{align}
where $\mathbf{U}_{N}^{(t_{j})} \in \mathbb{R}^{N}$ is a solution vector at time-step $t_j$.
The solution vector $\mathbf{U}_{N}^{(t_{j})}$ is high-dimensional, with $\mathrm{dim}\left(\mathbf{U}_N(\mathbf{X},t)\right)>>1$, making numerical computation of the solution prohibitively time consuming.

We consider the task of finding a reduced-order model of the system where $\mathrm{dim}\left(\mathbf{U}_r(\mathbf{X},t)\right)<< \mathrm{dim}\left(\mathbf{U}_N(\mathbf{X},t)\right)$ and making the solution of Eq. (\ref{eqn:PDE}) tractable.
The aim of this study is to build a combined space-time reduced-order model which takes into account the entire space-time cylinder $\Omega \times (0, T)$.
Such a reduced-order model that captures the coupled spatio-temporal correlation can provide an improved approximation of the solution field $\mathbf{U}(\mathbf{X}, t)$, which is dependent on both space $\mathbf{X}$ and time $t$.
In this study, we employ a 3D convolution network to learn a combined space-time reduced order model. 
The key details of the model are presented below.

\subsection{Combined space-time reduced-order model}
To learn a combined space-time reduced order model, we employ a 3D convolutional network.
First, a stack of convolutional and fully connected neural networks project the high-dimensional data to a reduced-dimension latent space.
Using a 3D convolution network, a mapping from the low-dimensional manifold to a high-dimensional space is subsequently learned, along with temporal evolution.
To capture the coupled spatio-temporal correlation, we apply 3D convolutional kernels to 2D fluid flow data, with the third dimension representing the time evolution of the fluid flow.
Backpropagation of error is used to update the weights of the 3D convolution kernel.
%
A 3D convolutional network based reduced order model can be employed to construct the approximation $\tilde{\mathbf{U}}_{N}^{(t)}$ using the full state solutions $ \mathbf{U}_{N}^{(t)}$ as follows:
\begin{subequations}
\begin{align}
\mathbf{U}_{r}^{(t)}&=\boldsymbol{\Psi}_{E}\left(\mathbf{U}_{N,}^{(t)};\theta_{E}\right),\\
\tilde{\mathbf{U}}_{N}^{(t+1)}&=\boldsymbol{\Psi}_{D}\left(\mathbf{U}_{r}^{(t+1)};\theta_{D}\right),
\end{align}
\end{subequations}
where $ \tilde{\mathbf{U}}_{N}^{(t+1)} \in \mathbb{R}^{N}$ denotes the approximation of the full state at time-step $t+1$, $\mathbf{U}_{N}^{(t)} \in \mathbb{R}^{N}$ represents the full state solution at time-step $t$,  $\boldsymbol{\Psi_{E}(.;\theta_{E})}$ stands for the encoder network that maps the full state to a low-dimensional manifold, $\theta_{E}$ is the parameter of encoder, $\boldsymbol{\Psi_{D}(.;\theta_{D})}$ denotes the decoder network that maps the low-dimensional data back to the high-dimensional physical state, and $\theta_{D}$ is the parameter of decoder. 
By backpropagating the $L_2$ norm of error, the weights of the network can be trained.
Here $ \mathbf{U}_{r}^{(t)} \in \mathbb{R}^{r}$ represents the solution on the reduced-dimension space. 

We briefly discuss a few of the convolutional network's distinguishing features that make it appropriate for a complete space-time reduced order modeling.
To begin, the solution of nonlinear time-dependent PDEs consists of the evolution of state variables from an initial condition, with an initial pattern propagating through the domain.
The translational invariance property of convolutional neural networks  aids in modelling this propagation of the disturbance through the domain.
Furthermore, a 3D convolution is appropriate for learning spatio-temporal features because it can better capture the combined spatial-temporal information compared to a 2D convolution.
In a 3D convolution model, convolution operations are performed spatio-temporal, but only spatially in a 2D convolution. Thus a 3D convolution preserves the temporal information of the input signals resulting in an output 3D volume. Similarly, for downsampling the data, we use 3D max pooling operations.
We only use max pooling in the spatial dimension to preserve information in the temporal dimension and keep the data output volume constant.

Convolutional neural network layers are often organized into feature maps, with each unit in a feature map linked to the local domain of a prior layer via a filter.
Consider the following three-dimensional input, $U \in \mathbb{R}^{N_t \times N_x \times N_y}$, where the first two dimensions represent spatial information and the third dimension represents temporal information. 
The convolutional layer for this input is composed of a collection of $F$ filters $K^f \in \mathbb{R}^{a \times b \times c}$ where each of these filters/kernels are three dimensional. 
These kernels/filters create a feature map $O^f \in \mathbb{R}^{n_t \times n_x \times n_y}$ through a three-dimensional discrete convolution, and non-linearity $\sigma$, which can be expressed as follows:
\begin{equation}
    \mathbf{O}_{i, j, k}^{f}=\sigma\left(\sum_{p} \sum_{q} \sum_{r}\mathbf{K}_{p, q, r}^{f} {U}_{(i-p),(j-q),(k-r)}\right).
\label{eqn:discrete_conv}
\end{equation}

%
For introducing a pointwise non-linearity, one can employ various activation functions such as sigmoid activation $\sigma(z) = \left(1+e^{- z }\right)^{-1}$.
The activation function allows nonlinear flow features to be captured in the discrete convolutional process. Furthermore, a pooling or down-sampling layer $\mathbf{g}=P(\mathbf{\mathbf{U}})$ may be used, which is given as
\begin{equation}
g_{l}(\mathbf{x})=P\left(\left\{\mathbf{O}_{l}\left(\mathbf{x}^{\prime}\right): \mathbf{x}^{\prime} \in \mathcal{N}_{e}(\mathbf{x})\right\}\right), l=1, \ldots, F,
\end{equation}
where $\mathcal{N}_{e}(\mathbf{x}) \subset \Omega_N$ is a neighborhood around the input variable $\mathbf{x}$ and $P$ is a pooling operation such as $L_{1}$, $L_{2}$ or $L_{\infty}$ norm. 
In the current work, the input variable has both spatial and temporal dimensions, and we only use the pooling operator along the spatial dimension, as previously mentioned.
A convolutional network is constructed by composing several convolutional and pooling layers, obtaining a generic compositional representation as follows:
\begin{equation}
{\mathbf{\theta_{CNN}}}(\mathbf{U})=\left({K^{(L)}} \cdots P \cdots \circ {K^{(2)}} \circ {K^{(1)}}\right)(\mathbf{U})
\end{equation}
where $\mathbf{\theta}_{CNN}=\left\{K^{(1)}, \ldots, K^{(L)}\right\}$ is the hyper-vector of the network parameters consisting of all the filter/kernel banks. 
Notably, Eq.~(\ref{eqn:discrete_conv}) is modified slightly if the convolutional filters/kernels are skipped on more than one element of the input signal along any Cartesian direction. The skipping lengths along the three directions of the input is termed as the stride $s_{L}=\left[\begin{array}{ll}s_{x} \;\;  s_{y} \;\; s_{z} \end{array}\right]$ and is an important hyperparameter for the dimensionality reduction. Convolutional neural networks possess multi-scale characteristics which allow them to scale easily to multi-dimensional Euclidean space \cite{bronstein2017geometric}.

In the present work, the 3D convolution network comprises an encoder and decoder path. The encoder path follows the typical architecture of a convolutional network. 
It consists of the repeated application of two $3\times3\times3$ convolutions (padded convolutions), each followed by
a batch normalization and Gaussian Error Linear Unit (GeLU).  
A $1\times2\times2$ max pooling operation with stride 1 is applied
for down-sampling the data in spatial dimensions. 
At certain down-sampling steps, we increase the number of filters
channels to increase the number of features. 
The decoder path consists of an up-sampling of the
feature map followed by  two $3\times3\times3$ convolutions (up-convolution), each followed by
a batch normalization and Gaussian Error Linear Unit (GeLU). 
At the final layer, a 3D convolution kernel is used with three feature channels to match the input data. 
In total,  the network has 25 convolutional layers.

\begin{figure*}
    \centering
    \includegraphics[]{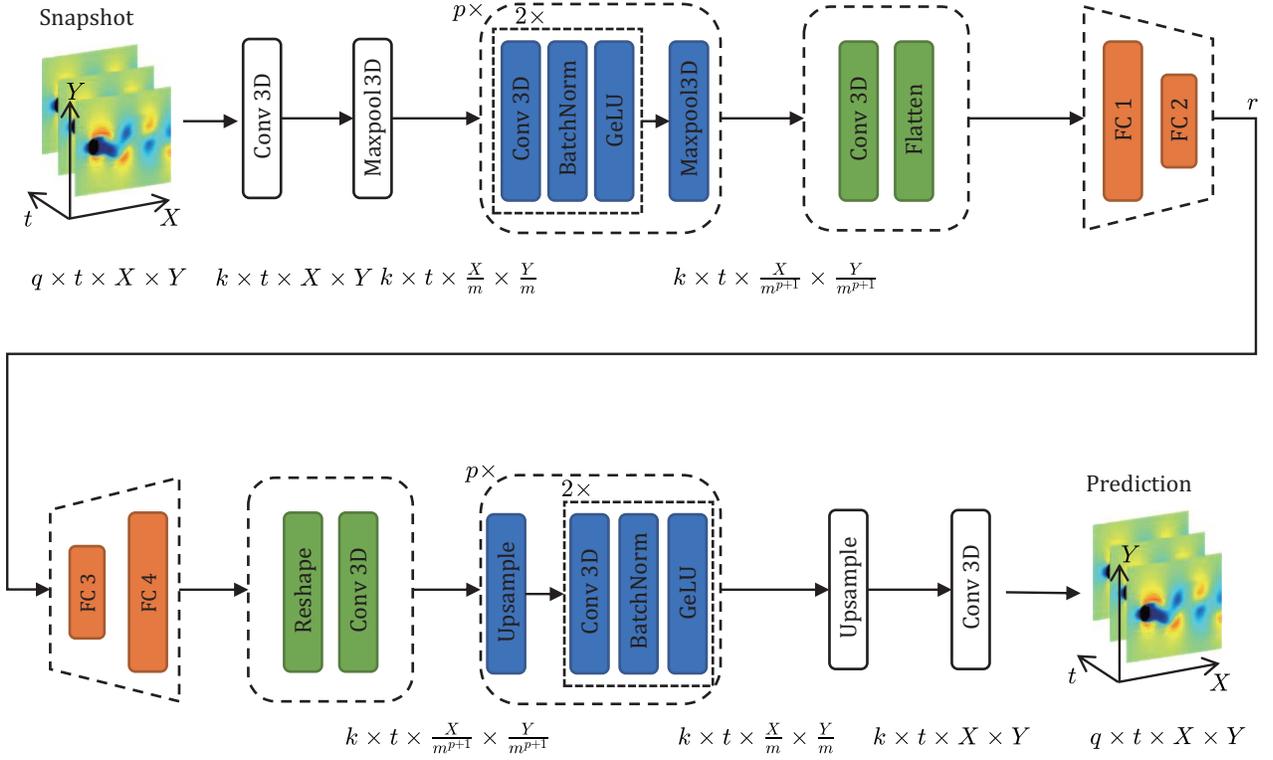}
    \caption{An illustration of the proposed 3D convolutional network. Here $q$ denotes the number of fluid flow variables, $k$ denotes the number of convolution filters, $p$ represents the number of computational blocks, $m$ stands for the size of maxpool kernel, and $r$ denotes the reduced dimension.}
\label{fig:3Dconv}
\end{figure*}

Training this 3D convolutional network then consists of finding the parameters that minimize the expected reconstruction error over all training examples given by
\begin{equation}
\boldsymbol{\theta}_{E}^{*}, \boldsymbol{\theta}_{D}^{*}=\arg \min _{\boldsymbol{\theta}_{\boldsymbol{E}}, \boldsymbol{\theta}_{D}}\mathcal{L}[\mathbf{U}_{N}^{(t+1)},\Psi_{D}(\Psi_{E}(\mathbf{U}_{N}^{(t)};\theta_{E});\theta_{D})],
\label{eqn: autoencoder loss}
\end{equation}
where $\mathcal{L}[\mathbf{U}_{N}^{(t+1)},\Psi_{D}(\Psi_{E}(\mathbf{U}_{N}^{(t)};\theta_{E});\theta_{D})]$ is a loss function in the $L^{2}$ norm,  which minimizes the difference between the reconstruction $\left(\Psi_{D}(\Psi_{E}(\mathbf{U}_{N}^{(t)};\theta_{E});\theta_{D})\right)$ and the ground truth $\left(\mathbf{U}_{N}^{(t+1)}\right)$.
A representative sketch of the proposed 3D convolutional network is shown in Fig.~\ref{fig:3Dconv}.


\section{Training strategy}
In this section, we further discuss the generation of the training dataset and training process.
For the training, we select 41 Reynolds numbers within the training Reynolds number range, leading to 41 separate runs to generate the data. The training set is sampled with a uniform interval in the Reynolds number range i.e., $\text{Re}=80,82,84,\ldots,160$. 
For each Reynolds number, we have 501 time steps of fluid flow data.
The flow data for each run is then decomposed into 482 pieces, with each piece containing 20 consecutive time steps, and then each of these pieces is divided into input (first 10 steps, ${X}_{\text{Train}}$) and output (last 10 steps, ${Y}_{\text{Train}}$) and serve as a single sample.

For improved neural network training and to prevent the over-saturation of any particular feature, the training data for each Reynolds number are scaled appropriately as follows: 
\begin{subequations}
\begin{align}
\overline{X}_{\text{Train}}&=\frac{{X}_{\text{Train}} - {X}_{\text{Train}_{,\min }}}{{X}_{\text{Train}_{,\max }}-{X}_{\text{Train}_{,\min }}},\\
\overline{Y}_{\text{Train}}&=\frac{{Y}_{\text{Train}} - {Y}_{\text{Train}_{,\min }}}{{Y}_{\text{Train}_{,\max }}-{Y}_{\text{Train}_{,\min }}},
\end{align}
\end{subequations}
where $\overline{X}_{\text{Train}}, \overline{Y}_{\text{Train}}  \in[0,1]^{N_m\times N_t \times N_x \times N_y}$ are min-max normalized training data. This training set is used for training the model. 
\begin{figure*}
  \centering
  \includegraphics[]{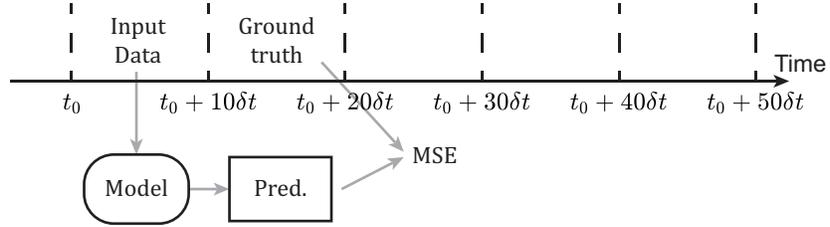}  
\caption{Schematic of the model training: Input and output from model, and calculation of mean squared error loss.}
\label{fig:fig_training_val}
\end{figure*}

The 3D convolution network makes predictions in a sequence-to-sequence manner: At time step $t_0$, the network is provided with the state history of the flow field (velocity and pressure) in the last $N_h=10$ time steps, and expected to predict the state of the flow in the next non-overlapping $N_p=10$ time steps. 
The input dimensions are $3 \times 10 \times 128 \times 128$.
It is assumed that the Reynolds number is constant over these time steps and that the flow has reached a steady state over the whole domain for all the training and evaluation data.
The training procedure is illustrated in Fig. \ref{fig:fig_training_val}.

We want to determine the model parameters  $\theta^* = \left\{\theta_E^*,\theta_D^*\right\}$ such that they minimize the following expected error between the model prediction and the ground truth data 
\begin{equation}
\begin{aligned}
\mathcal{L}\left(\hat{{X}}, {Y}\right)
=\frac{1}{N_m}\sum_{j=1}^{N_m}\left( \frac{1}{N_t}\sum_{i=1}^{N_{T}} \left\|\hat{X}_{{Re}_j}^{i}-{Y}_{ {Re}_j}^{i}\right\|_{2}^{2} \right),
\end{aligned}
\label{eqn:loss}
\end{equation}
where $\hat{{X}}$ and ${Y}$ are the model prediction and the ground truth respectively. While $N_m$ denotes the numbers of Reynolds number, $N_t$ represents the time steps for each Reynolds number.


In the present work, we utilize a step-based learning rate scheduler which reduces the initial learning rate by a decay factor on every predetermined number of steps \cite{bengio2012practical}.
We use the AdamW optimizer \cite{loshchilov2018decoupled}, a version of adaptive gradient algorithm Adam.
Algorithm \ref{alg:alg1} provides a complete training procedure for our architecture.

\SetKwInput{KwOutput}{Output}

\begin{algorithm}
\caption{3D Convolution Training Algorithm}
\KwIn{${X}_{\text{Train}}$, ${Y}_{\text{Train}}$, $N_{\text{epochs}}$}
\KwOutput{$\{\theta^{*}  = \theta_{E}^{*},\theta_{D}^{*}\}$}
 Initialize $\theta$\;
 \While{ epoch $<$ $N_{\text{epochs}}$ }
 {
 Randomly sample batch from training data: ${X}_{\text{Train}}^{b} \subset {X}_{\text{Train}}$\;
 Forward pass: ${{\hat{X}}}^{b} \leftarrow \Psi_{D}\left(\Psi_{E}\left({{X}}^{b};\theta_{E}\right);\theta_{D}\right)$, ${\hat{X}}^{b} \in \mathbb{R}^{N_b \times N_t \times N_x \times N_y}$\;
Calculate loss $\mathcal{L}$ via Eq.~(\ref{eqn:loss})\;
Estimate gradients $\hat{\mathbf{g}}$ using automatic differentiation\;
Update parameters: $\theta \leftarrow A D A M W(\hat{\mathbf{g}})$\;
}
Updated parameters: $\{\theta^{*}  = \theta_{E}^{*},\theta_{D}^{*}\}$
\label{alg:alg1}
\end{algorithm}

The process of prediction becomes relatively simple once the model has been trained. The network is employed to predict the next sequence of time steps using the input sequence ${X}_{\text{in}}$ and the trained parameters $\theta^{*}$. The predicted time steps are then evolved for $n$ time-horizons ($N_{th}$) by an iterative application of the network for generating the whole temporal sequence of data. Algorithm \ref{alg:alg2} explains how to make a prediction using our proposed framework.
\begin{algorithm}
\caption{3D Conv Prediction Algorithm}
\KwIn{${X}_{\text{in}}$, $N_{th}$}
\KwResult{Model prediction $\hat{X}_{\text{out}}$ }
 Load trained parameter $\theta^{*} =\{\theta_{E}^{*},\theta_{D}^{*}\}$\;
 \While{ i $<$ $N_{\text{th}}$ }
 {
 Forward pass: ${{\hat{X}}} \leftarrow \Psi_{D}\left(\Psi_{E}\left({{X}};\theta_{E}\right);\theta_{D}\right)$\;

Append: $\hat{X}_{\text{out}} \leftarrow {{\hat{X}}}$\;
}

Output: $\hat{X}_{\text{out}}$
\label{alg:alg2}
\end{algorithm}

The testing data sets are generated with $\text{Re} \in [20,300]$ with the increment of 20, and processed in the same way as the training set. The extrapolation capability of the network is evaluated via the MSE loss between the network output ($\hat{{X}}$) on these evaluation data sets and the corresponding ground truth (${Y}$). The MSE loss is given by:
\begin{equation}
\operatorname{MSE}(\hat{X}, {{Y}})=\sum_{i=1}^{N}\sum_{j=1}^{N}\frac{(\hat{{X}}_{i,j}^t-{Y}_{i,j}^t)^{2}}{N},
\end{equation}
 where $i$ and $j$ are indices for the location on the spatial grid, superscript $t$ denotes the time-step, $N$ is the number of spatial grid points.
We compute the temporal average for every Reynolds number to obtain a single metric for comparison across different Reynolds numbers. The MSE formula for the temporal average as a function of the Reynolds number is given by:
 \begin{equation}
<\operatorname{MSE}(\hat{X}, {{Y}})(Re)>=\sum_{t=1}^{N_t}\frac{\left( \sum_{i=1}^{N}\sum_{j=1}^{N}\frac{(\hat{{X}}_{i,j}^t-{Y}_{i,j}^t)^{2}}{N} \right)}{N_t},
\label{eq:temp_avg}
\end{equation}
where $N_t$ is number of time steps.

\begin{figure*}
  \centering
  \includegraphics[]{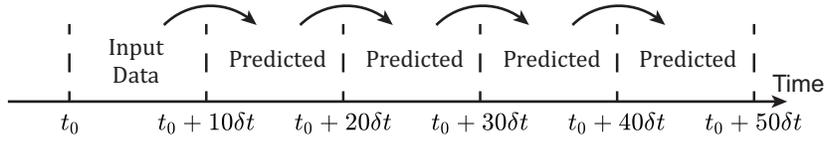}  
\caption{Illustration of predicting the temporal series for the evaluation parameter.}
\label{fig:fig_val}
\end{figure*}

We evaluate our model on the testing data set. 
We start with the first ten time steps and then auto-regressively predict the next ten. 
In the present method, we iterate 50 more times into the future, predicting up to 500 time steps from the start. 
This is in contrast to our training scenario, where the loss is computed only from one iteration.
Fig. \ref{fig:fig_val} shows the evaluation procedure. 
We calculate the temporal average for each time series data using Eq. \ref{eq:temp_avg}.
We evaluate the performance of various models in the extrapolation regime of the Reynolds number using the temporal-average mean squared error.
Next, we turn our attention to a detailed assessment of the models.

\section{Numerical Results}
In this section, we demonstrate how the proposed architecture can predict the evolution of fluid flow and its performance in the extrapolation regime of the Reynolds number. The reliability and accuracy of our proposed methodology will be demonstrated for a test problem of the flow past a circular cylinder.
We compare our results against the convolutional recurrent autoencoder network \cite{bukka2020deep,bukka2021assessment}

\subsection{Flow past a circular cylinder}
As a test case, let us consider the flow over a 2D circular cylinder. 
The fluid flow is given by the incompressible Navier-Stokes equations:
\begin{subequations}
\begin{align}
\nabla u^{*}(\mathbf{X},t) &= 0, \\
\frac{\partial u^{*}(\mathbf{X},t)}{\partial t^{*}}+\left(u^{*}(\mathbf{X},t) \cdot \nabla\right) u^{*}(\mathbf{X},t)&=-\nabla p^{*}+\frac{\mathbf{1}}{\boldsymbol{R} \boldsymbol{e}} \Delta^{*} u^{*}(\mathbf{X},t),    
\end{align}
\end{subequations}
where ${u}^{*}({x}, t)$ and $p^{*}({x}, t)$ are the non-dimensional velocity and pressure fields, respectively. The Reynolds number is $R e=U D / \nu$, where $\nu$ is the constant kinematic fluid viscosity, $D$ is the cylinder diameter, and $U$ represents the free-stream  velocity. We consider a fixed-size domain in a 2D plane.
Given a specific geometry fixed within the domain, we assume that sufficient flow data is sampled from the fluid flow past the geometry within the Reynolds number range $\text{Re}\in[\text{Re}_{min},\text{Re}_{max}]$.
 In this work, we select the training Reynolds number range $\text{Re}_{tr}\in[80,160]$, and evaluate the extrapolation error in the Reynolds number range $\text{Re}_{ex}\in[20,80)\cup(160,300]$.

 \begin{figure*}
  \centering
  \includegraphics[]{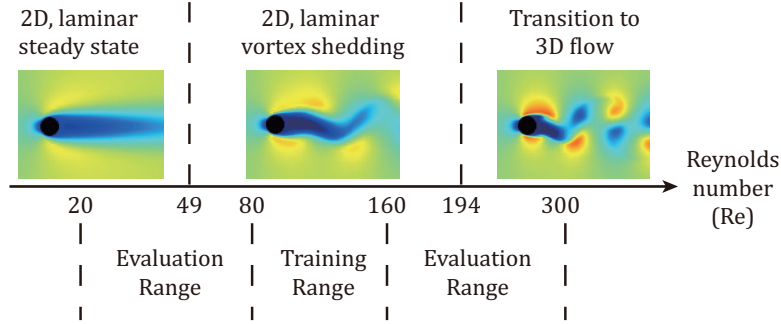}  
\caption{Illustrating different regimes of fluid flow pattern in the wake of the cylinder for different values of Reynolds number, and  the training and evaluation range of Reynolds numbers.}
\label{fig:fig_training}
\end{figure*}
 
The flow data used for the training and evaluation is generated with the open-source CFD framework Basilisk \cite{Popinet2013}.  Figure \ref{fig:fig_training} shows the fluid flow over the cylinder with varying Reynolds numbers.
The inlet velocity is fixed at $U_{x,in}=1$, whilst the Reynolds number is controlled by varying the kinematic viscosity $\nu$. The simulation runs up to time $t=15$ on an adaptive grid with maximum resolution $512\times512$. For each run, the simulated flow data is sampled from time step $t_{start}=10$ to $t_{end}=15$ with time interval $\Delta t=0.01$ on a uniform $257\times257$ grid. The flow data before $t_{start}$ are not used so that only flow in a steady state is considered. We utilize a bilinear interpolation for the regions where the grid is relatively coarse and no exact flow data on the sampling point is available. Only a part within the sampled flow field of size $128\times128$ is employed for the training and testing.

The flow around a smooth circular cylinder strongly depends on the Reynolds number. As the Reynolds number is increased from zero, the flow patterns vary dramatically. For a very low Reynolds number, the flow tends to be laminar and there is no separation \cite{sumer2006hydrodynamics} and the fluid wraps around the cylinder. For the Reynolds number range of $5<Re<45$, the flow becomes separated and a pair of steady vortices are observed in the wake of cylinder \cite{wiliamson1996vortex}. 
For the Reynolds number range of $45<Re<200$, the flow becomes unstable and sheds alternate opposite-signed vortices periodically which is known as a Karman vortex street \cite{williamson1989oblique}. At a higher Reynolds number of $Re>300$, the vortex wake behind the cylinder becomes turbulent and chaotic. A neural network trained with flow data from any of these regimes should be able to predict the flow evolution not only for any Reynolds number within the range (interpolation) but also for any Reynolds number outside of the selected range (extrapolation). While a perfect extrapolation may be challenging in practice, we strive for a small extrapolation error near the training Reynolds number range.

The architecture used in this test case is described as follows. 
We choose a 25-layer convolutional layer. The encoder consists of a 3D convolutional layer, max pooling and fully connected layers. There are {$r$} neurons in the output layer of the encoder function. Specific details of the encoder and the decoder functions are summarized in Table \ref{tab:LC_ENC}, and \ref{tab:LC_Dec}. 
The total number of trainable parameters (i.e., weights and biases) of the neural network for this case is 1,103,827. 
{The model was trained from scratch with PyTorch \cite{DBLP:journals/corr/abs-1912-01703} using a single NVIDIA V100 Tesla graphical processing unit and trained for 30 epochs and 8 hours of wall clock time.}

\begin{table*}
\caption{Attributes of the encoder $\boldsymbol{\Psi}_{E}(.;\theta_E)$.}
\centering
\begin{ruledtabular}
\begin{tabular}{lllllll}
\text { Layer } & $\begin{array}{l}
\text { Layer } \\
\text { Type }
\end{array}$& $\begin{array}{l}
\text { Input } \\
\text { Dimension }
\end{array} $& $\begin{array}{l}
\text { Output } \\
\text { Dimension }
\end{array} $& $\begin{array}{l}
\text { Kernel } \\
\text { Size }
\end{array} $& $\begin{array}{l}
\text { \# filters/ } \\
\text { \# neurons }
\end{array} $& \text { Stride }\\
\hline 1 & \text{Conv 3D} & {[3,10,128,128]} & {[16,10,128,128]} & {[3,3,3]} & 16 & [1,1,1] \\
 & \text{MaxPool 3D} &{[16,10,128,128]} & {[16,10,64,64]} & {[1,2,2]} & - & - \\
 2 & \text{Conv 3D} & {[16,10,64,64]} & {[16,10,64,64]} & {[3,3,3]} & 16 & [1,1,1] \\
  3 & \text{Conv 3D} & {[16,10,64,64]} & {[32,10,64,64]} & {[3,3,3]} & 32 & [1,1,1] \\
  & \text{MaxPool 3D} &{[32,10,64,64]} & {[32,10,32,32]} & {[1,2,2]} & - & - \\
4 & \text{Conv 3D} & {[32,10,32,32]} & {[32,10,32,32]} & {[3,3,3]} & 32 & [1,1,1] \\
5 & \text{Conv 3D} & {[32,10,32,32]} & {[32,10,32,32]} & {[3,3,3]} & 32 & [1,1,1] \\
& \text{MaxPool 3D} &{[32,10,32,32]} & {[32,10,16,16]} & {[1,2,2]} & - & - \\
6 & \text{Conv 3D} & {[32,10,16,16]} & {[32,10,16,16]} & {[3,3,3]} & 32 & [1,1,1] \\
7 & \text{Conv 3D} & {[32,10,16,16]} & {[32,10,16,16]} & {[3,3,3]} & 32 & [1,1,1] \\
& \text{MaxPool 3D} &{[32,10,16,16]} & {[32,10,8,8]} & {[1,2,2]} & - & - \\
8 & \text{Conv 3D} & {[32,10,8,8]} & {[32,10,8,8]} & {[3,3,3]} & 32 & [1,1,1] \\
9 & \text{Conv 3D} & {[32,10,8,8]} & {[32,10,8,8]} & {[3,3,3]} & 32 & [1,1,1] \\
& \text{MaxPool 3D} &{[32,10,8,8]} & {[32,10,4,4]} & {[1,2,2]} & - & - \\
10 & \text{Conv 3D} & {[32,10,4,4]} & {[32,10,4,4]} & {[3,3,3]} & 32 & [1,1,1] \\
11 & \text{Conv 3D} & {[32,10,4,4]} & {[64,10,4,4]} & {[3,3,3]} & 64 & [1,1,1] \\
& \text{MaxPool 3D} &{[64,10,4,4]} & {[64,10,2,2]} & {[1,2,2]} & - & - \\
12 & \text{Conv 3D} & {[64,10,2,2]} & {[64,10,2,2]} & {[3,3,3]} & 64 & [1,1,1] \\
 & \text{Flatten} & {[64,10,2,2]} & {[2560]} & - & - & - \\
 & \text{Dense} &{[2560]} & {[64]} & -  & 64 & -\\
 & \text{Dense} &{[64]} & {[r]} & -  & r & -\\
\end{tabular}

  \label{tab:LC_ENC}
\end{ruledtabular}
\end{table*}

\begin{table*}
\caption{Attributes of the decoder $\boldsymbol{\Psi}_{D}(.;\theta_D)$.}
\centering
\begin{ruledtabular}
\begin{tabular}{lllllll}
\text { Layer } & $\begin{array}{l}
\text { Layer } \\
\text { Type }
\end{array}$& $\begin{array}{l}
\text { Input } \\
\text { Dimension }
\end{array} $& $\begin{array}{l}
\text { Output } \\
\text { Dimension }
\end{array} $& $\begin{array}{l}
\text { Kernel } \\
\text { Size }
\end{array} $& $\begin{array}{l}
\text { \# filters/ } \\
\text { \# neurons }
\end{array} $& \text { Stride }\\
\hline 
 & \text{Dense} &{[r]} & {[64]} & -  & 64 & -\\
 & \text{Dense} &{[64]} & {[2560]} & -  & 2560 & -\\
 & \text{reshape} &{[2560]} & {[64,10,2,2]} & -  & - & -\\
13 & \text{Conv 3D} & {[64,10,2,2]} & {[64,10,2,2]} & {[3,3,3]} & 64 & [1,1,1] \\
 & \text{Upsample} &{[64,10,2,2]} & {[64,10,4,4]} & {[1,2,2]} & - & - \\
 14 & \text{Conv 3D} & {[64,10,4,4]} & {[32,10,4,4]} & {[3,3,3]} & 32 & [1,1,1] \\
15 & \text{Conv 3D} & {[32,10,4,4]} & {[32,10,4,4]} & {[3,3,3]} & 32 & [1,1,1] \\
 & \text{Upsample} &{[32,10,4,4]} & {[32,10,8,8]} & {[1,2,2]} & - & - \\
16 & \text{Conv 3D} & {[32,10,8,8]} & {[32,10,8,8]} & {[3,3,3]} & 32 & [1,1,1] \\
17 & \text{Conv 3D} & {[32,10,8,8]} & {[32,10,8,8]} & {[3,3,3]} & 32 & [1,1,1] \\
& \text{Upsample} &{[32,10,8,8]} & {[32,10,16,16]} & {[1,2,2]} & - & - \\
18 & \text{Conv 3D} & {[32,10,16,16]} & {[32,10,16,16]} & {[3,3,3]} & 32 & [1,1,1] \\
19 & \text{Conv 3D} & {[32,10,16,16]} & {[32,10,16,16]} & {[3,3,3]} & 32 & [1,1,1] \\
& \text{Upsample} &{[32,10,16,16]} & {[32,10,32,32]} & {[1,2,2]} & - & - \\
20 & \text{Conv 3D} & {[32,10,32,32]} & {[32,10,32,32]} & {[3,3,3]} & 32 & [1,1,1] \\
21 & \text{Conv 3D} & {[32,10,32,32]} & {[32,10,32,32]} & {[3,3,3]} & 32 & [1,1,1]\\
& \text{Upsample} &{[32,10,32,32]} & {[32,10,64,64]} & {[1,2,2]} & - & - \\
22 & \text{Conv 3D} & {[32,10,64,64]} & {[16,10,64,64]} & {[3,3,3]} & 16 & [1,1,1]\\
23 & \text{Conv 3D} & {[16,10,64,64]} & {[16,10,64,64]} & {[3,3,3]} & 16 & [1,1,1]\\
& \text{Upsample} & {[16,10,64,64]} &  {[16,10,128,128]} & {[1,2,2]} & - & - \\
24 & \text{Conv 3D} & {[16,10,128,128]} & {[16,10,128,128]} & {[3,3,3]} & 16 & [1,1,1]\\
25 & \text{Conv 3D} & {[16,10,128,128]} & {[3,10,128,128]} & {[3,3,3]} & 3 & [1,1,1]\\
\end{tabular}

  \label{tab:LC_Dec}
\end{ruledtabular}
\end{table*}

\subsection{Prediction from 3D Convolution}
\begin{figure*}
  \centering
\includegraphics[]{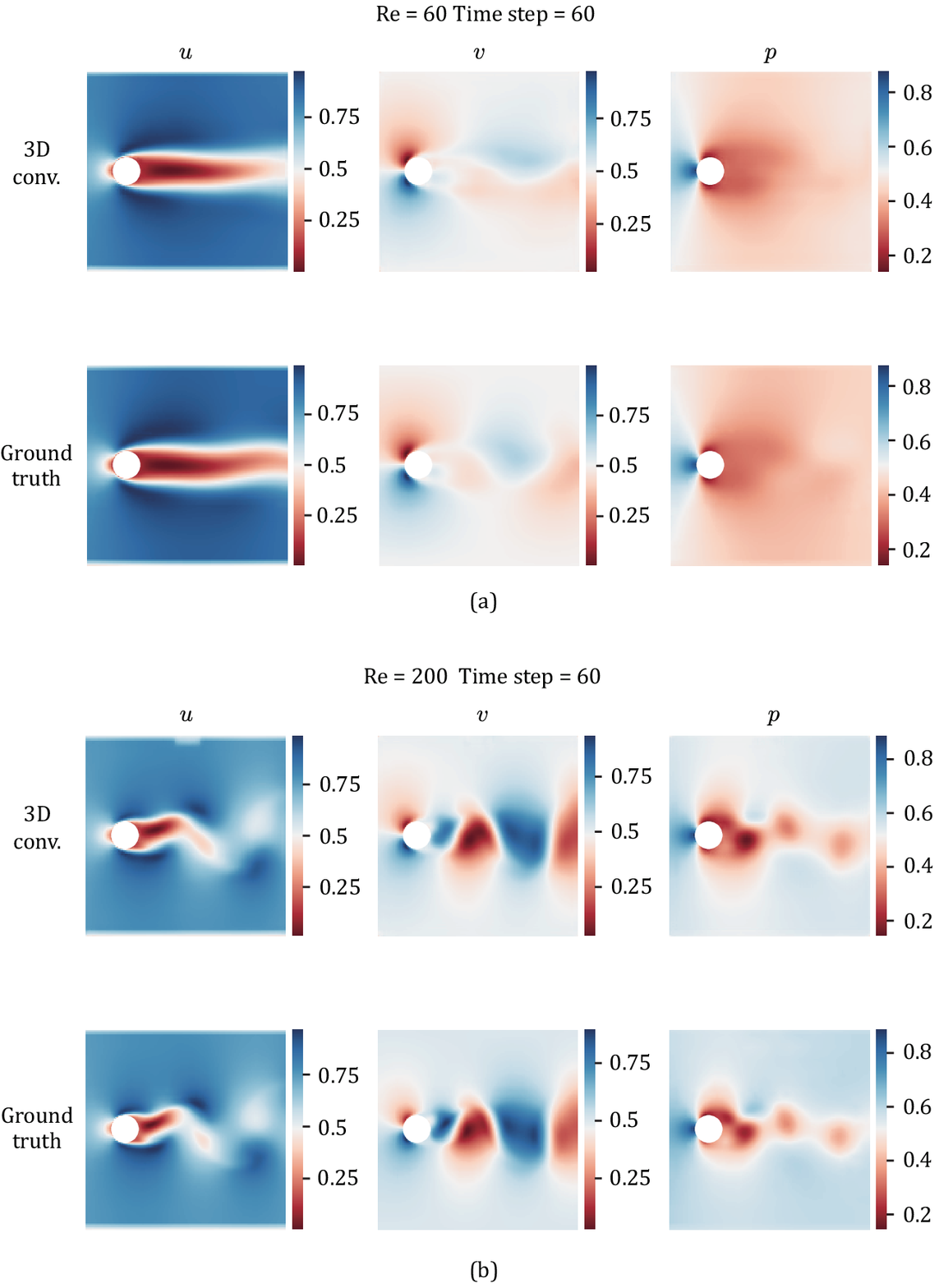}
\caption{Ground truth and prediction from 3D convolution architecture at $60^{\text{th}}$ time-step: (a) Flow field for Re = 60, (b) flow field for Re = 200.}
\label{fig:fig_TotConv_Prediction}
\end{figure*}

In this section, we will assess the predictions of our 3D convolution framework for varying Reynolds numbers in the extrapolation regime. In particular, we will consider two test cases, namely: Test Case 1, corresponding to a Reynolds number of 60 and Test Case 2 with a Reynolds number of 200.
The first ten time steps of data from the incompressible Navier-Stokes equation with $Re = 60$ are fed into the model, which predicts auto-regressively for up to 500 timesteps. 
Figure~ \ref{fig:fig_TotConv_Prediction}a illustrates both the exact solution and the 3D convolution approximation for this instance of the testing parameter.
The laminar flow phenomenon of low Reynolds number flow is accurately captured by the 3D convolution architecture.

The fluid flow with $Re = 200$ corresponds to the vortex shedding and the transition to 3D flow. The Reynolds number for 2D data considered in this test case is similar to the training Reynolds number.
Similarly, the first ten time steps of data are fed into the model, which predicts up to 500 time steps auto-regressively.
The exact solution and the 3D convolution approximation for this instance of the testing parameter are shown in Fig. \ref{fig:fig_TotConv_Prediction}b. 
Because the physics of this flow corresponds to the training regime, the 3D convolution architecture accurately predicts the flow field for this test case as well. 
The vortex shedding phenomenon is accurately captured by the model.
The 3D Convolution architecture predicts the fluid flow accurately in the extrapolation regime close to the training data range. The temporal averaged mean squared error for the prediction from 3D convolution network for $Re$ = 60 and 200 are $1.325\times10^{-3}$ and $1.734\times10^{-3}$, respectively.



\subsection{Effect of network architectures}
In this section, we explore the sensitivity of various hyperparameters in the 3D convolution architecture and assess their impact on the predictive performance.
For this study, we are particularly interested in how deep networks can aggregate coupled spatio-temporal information.  To determine a good 3D convolution architecture, we change the number of convolution and max-pooling layers in the architecture.
We maintain the consistency of the input and output dimensions of the convolution layer while keeping zero padding and a stride of one. A $3 \times 3 \times 3$ convolution kernel is considered in this work.
Following each convolution layer, a 3D max pooling operation is carried out. We use a max pooling kernel of size $1 \times 2 \times 2$ which reduces the dimension only along the spatial dimension. The number of the maximum pooling operations determines the reduced dimension after the convolution encoder.
We experiment with $[5,6, 7]$ max pooling layers that lead to reduced dimensions of $4\times4$, $2\times2$, and $1\times1$ along the spatial dimension. Another important hyperparameter is the number of neurons, $r$, in the last full-connected layer, which determines the dimension of the reduced manifold. We select a reduced dimension of $[16,32,64]$ in our experiments.
The optimal hyper-parameter selection for these variables is achieved via grid search techniques provided by the ray tune library \cite{liaw2018tune}.

\begin{figure*}
  \centering
  \includegraphics[]{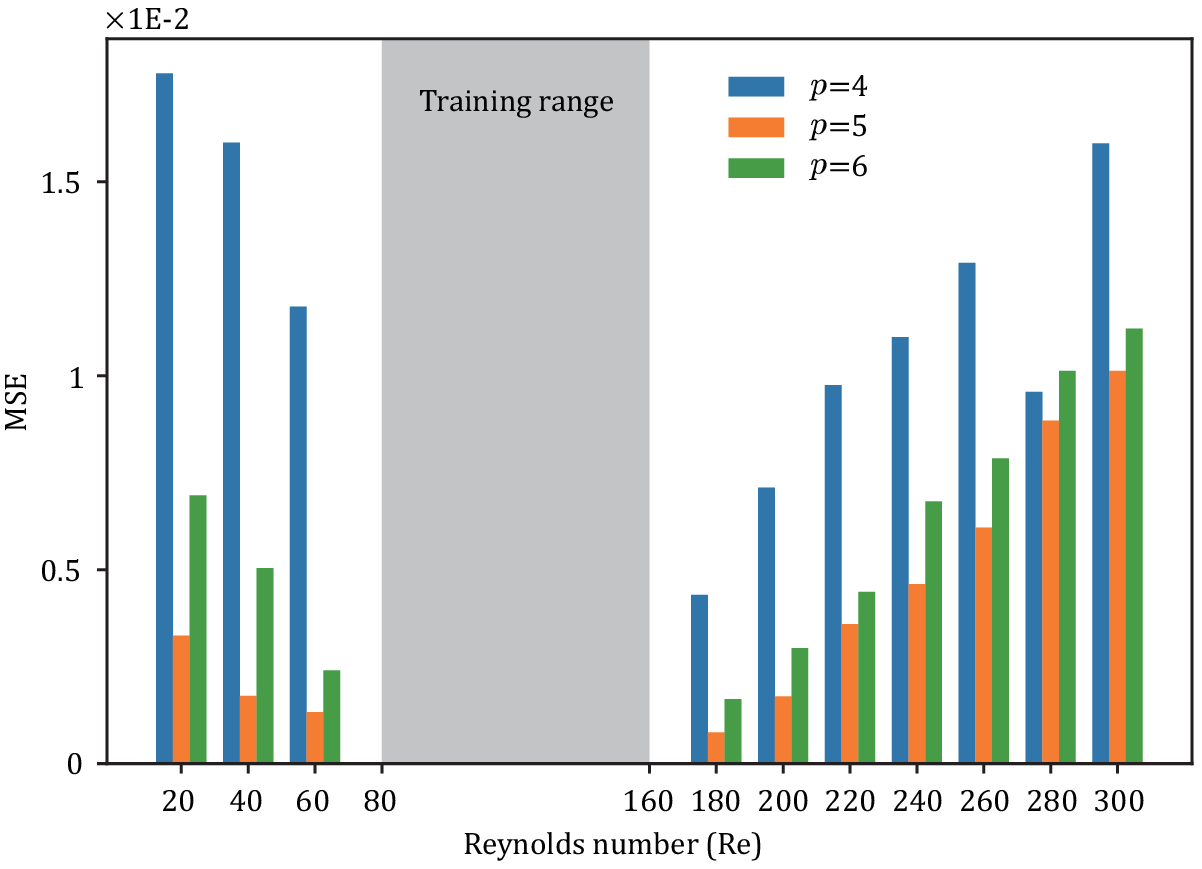}  
\caption{Dependence of the temporal averaged MSE on the Reynolds number in the extrapolation regime for 3D Conv. 1, 2, and 3, which correspond to the number of maxpool layers 5, 6, and 7.}
\label{fig:mse_TotalConv}
\end{figure*}

The mean squared errors of the various 3D convolution predictions are compared in Fig. \ref{fig:mse_TotalConv}.  
In the extrapolation regime, the 3D convolution with the 6 max-pooling layers and reduced dimension = 16 produces the lowest mean squared error for varying Reynolds numbers. Hence the optimal hyperparameter for 3D convolution is 6 max-pooling layers and reduced dimension of 16.

\subsection{Comparison with CRAN}
In this section, we compare the proposed 3D convolution model against the CRAN model. The CRAN architecture employs an encode-propagate-decode procedure for learning fluid dynamics. To implement this methodology, the CRAN architecture utilizes two forward passes, an autoencoder pass and a propagator pass. These two passes result in two loss functions, autoencoder loss and propagator loss which are combined using a hyperparameter. 
We employ the ray tune library to find the optimal value of this hyperparameter for the flow past a cylinder.
We contrast the CRAN's performance to that of our 3D convolution model, which learns coupled spatio-temporal correlation.
When compared to our 3D convolution model, the CRAN model uses 2D convolution to capture spatial correlation and long short-term memory cells to learn the temporal dynamics.
Further details about the CRAN architecture can be found in \cite{deo2022predicting,bukka2020deep}. 

\begin{figure*}
\centering
\includegraphics[]{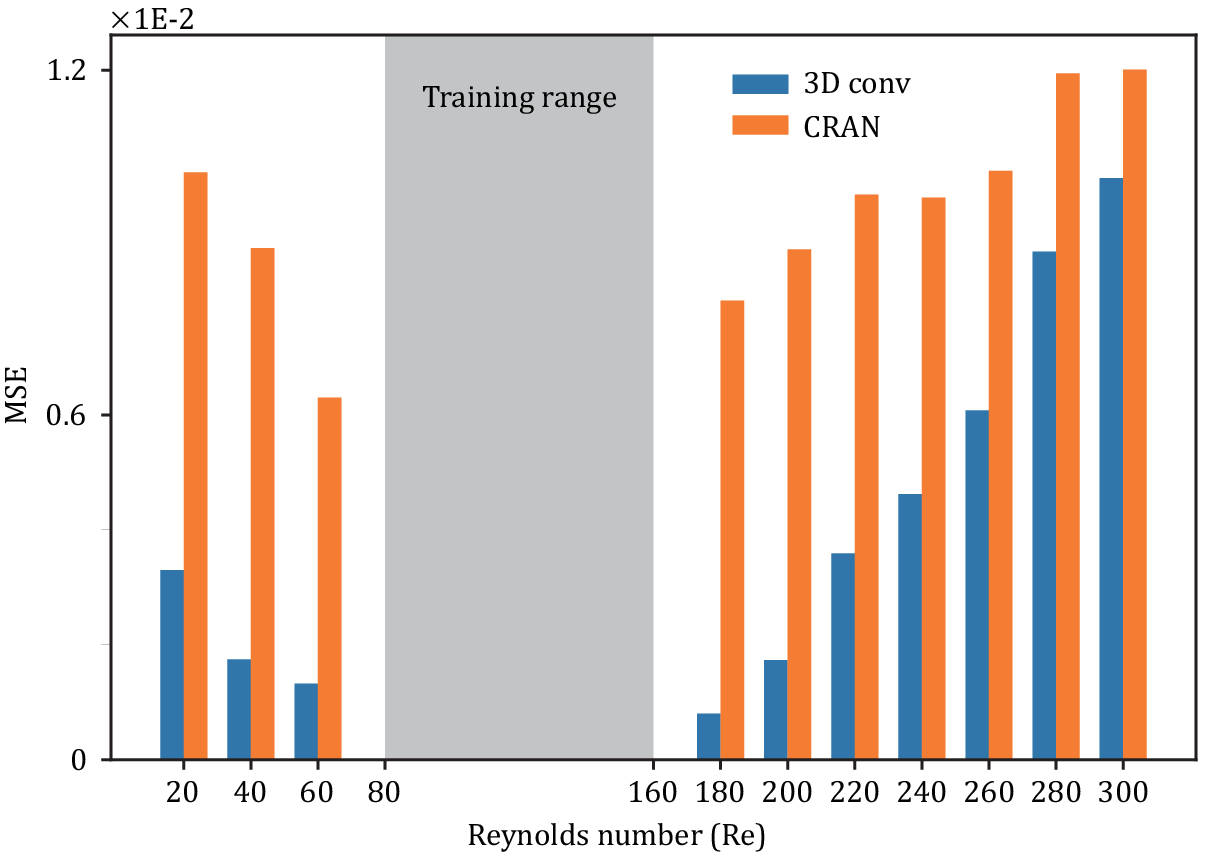}
 \caption{Temporal averaged MSE vs Reynolds number for CRAN and 3D convolution.}
\label{fig:MSE_CRAN_TotConv}
\end{figure*}

To assess the extrapolation capability,  we plot the temporal averaged MSE for the CRAN and the 3D convolution models with varying Reynolds numbers in Fig. \ref{fig:MSE_CRAN_TotConv}.
We can see that the loss grows monotonically in the out of the training range for the two models as expected because these Reynolds numbers correspond to different regime of the fluid flow.
It is worth noting that close to the training range the 3D convolution model performs better than the CRAN architecture, as shown in Fig. \ref{fig:MSE_CRAN_TotConv}. 
The result from the 3D convolution model shown by the blue bar plot outperforms the CRAN model in the low Reynolds number regime.
For a high Reynolds number regime, when the Reynolds number is close to the training regime, the 3D convolution model shows an improved performance than the CRAN architecture. 
These results validate our claim that infusing coupled spatial and temporal correlation as an implicit helps generalize better in the parameter space.

\begin{figure*}
    \centering
    \includegraphics[]{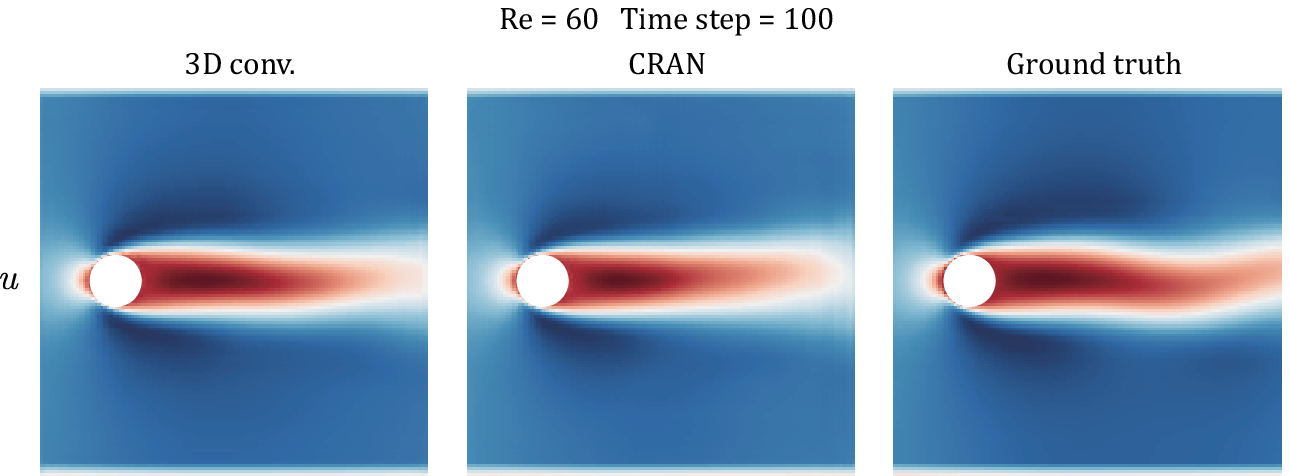}
    \caption{Flow over cylinder for Re = 60 at time-step = 100. From left to right, 3D convolution, CRAN and ground truth.}
\label{fig:Comparison}
\end{figure*}

\begin{figure*}
    \centering
    \includegraphics[]{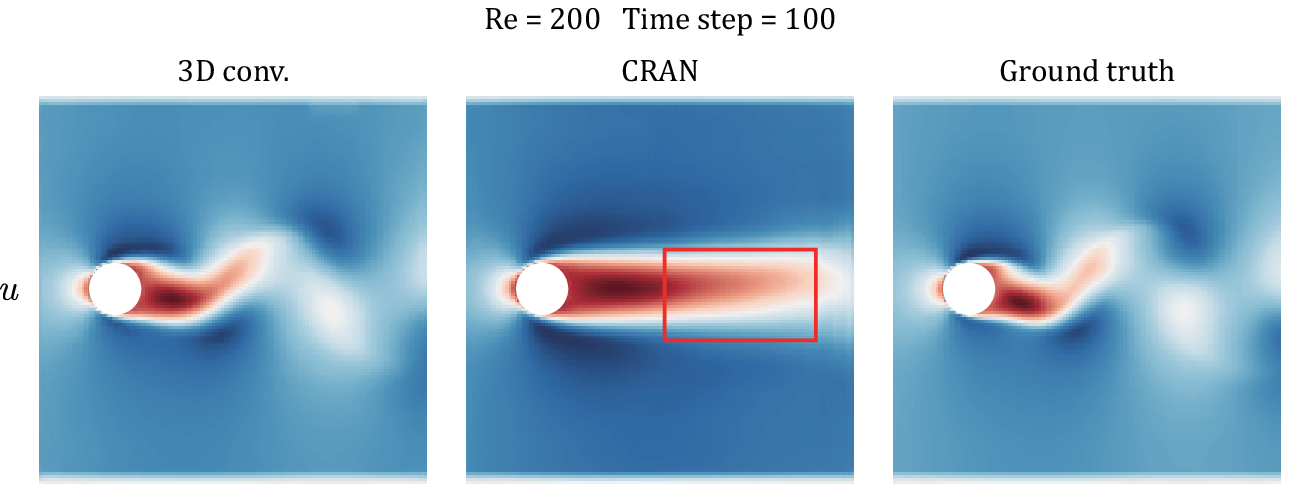}
    \caption{Flow over cylinder for Re = 200 at time-step = 100. From left to right, 3D convolution, CRAN and ground truth.}
\label{fig:Comparison2}
\end{figure*}

To better understand and analyze these results, we plot the predicted $u$-velocity component for the extrapolation case of  Reynolds number 60 in Fig. \ref{fig:Comparison}. 
As previously stated, there is no Karman vortex street at this Reynolds number, only a stable flow. 
Both the CRAN and 3D convolution models predict the stable flow behind the cylinder as shown in Fig. \ref{fig:Comparison}.
To further analyze these results, we plot the predicted $u$-velocity component for the extrapolation case of  Reynolds number 200 in Fig. \ref{fig:Comparison2}.
The flow field from the CRAN qualitatively looks the same as $Re$ = 60. 
These findings lead us to the conclusion that CRAN architecture is overfitting on the training data, which prompts us to reduce the number of trainable parameters in the CRAN architecture.

The expressivity and capacity of deep neural networks to approximate a function depend on the number of trainable parameters.
A deep neural network with a very large number of trainable parameters tends to overfit the training data.
To address the overfitting issue in CRAN, we reduced the number of trainable parameters in the CRAN architecture and termed the new architecture CRAN reduced (CRAN\_r). 
The total number of parameters in the CRAN reduced is around three hundred thousand compared to approximately 1.1 million parameters in 3D convolution and 1.7 million parameters in CRAN.

\begin{figure*}
    \centering
    \includegraphics[]{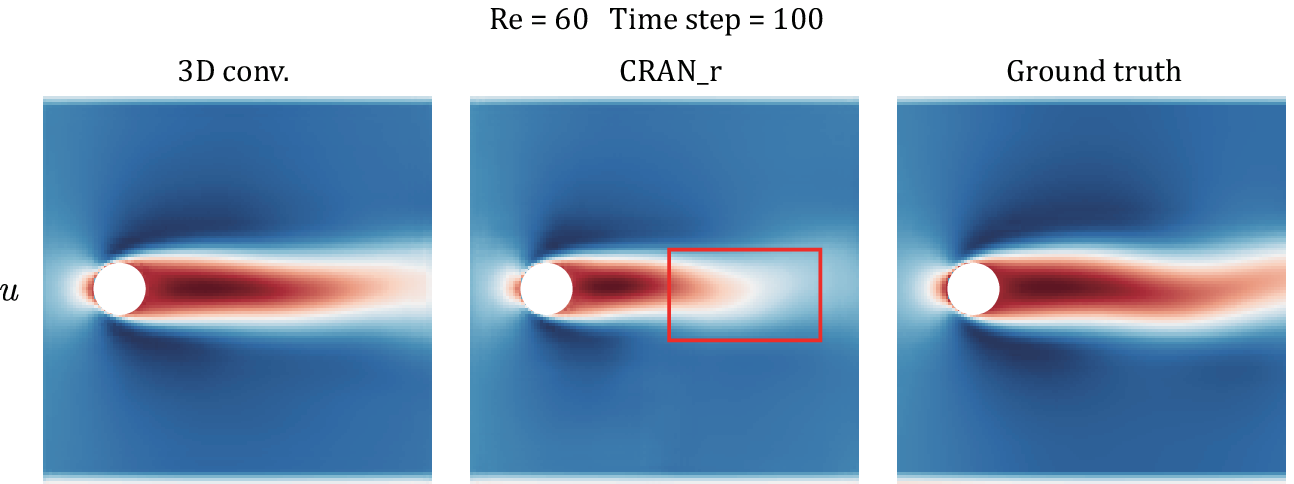}
    \caption{Flow over cylinder for $Re$ = 60 at time-step = 100. From left to right, 3D convolution, CRAN reduced and ground truth.}
\label{fig:Comparison3}
\end{figure*}

In Fig. \ref{fig:Comparison3}, we plot the predicted $u$ velocity from the updated CRAN model. 
The optimized CRAN with the reduced number of parameters fails to predict the wake behind the cylinder for the Reynolds number of $Re=60$, as seen in the red box in Fig. \ref{fig:Comparison3}. 
The 3D convolution architecture trained on the vortex flow regime is able to predict the wake but shows slight oscillation at the tail end.
To understand the extrapolation behaviour of 3D convolution, CRAN and CRAN\_r, we plot the temporal averaged MSE for Reynolds number outside of the training range.

\begin{figure*}
    \centering
    \includegraphics[]{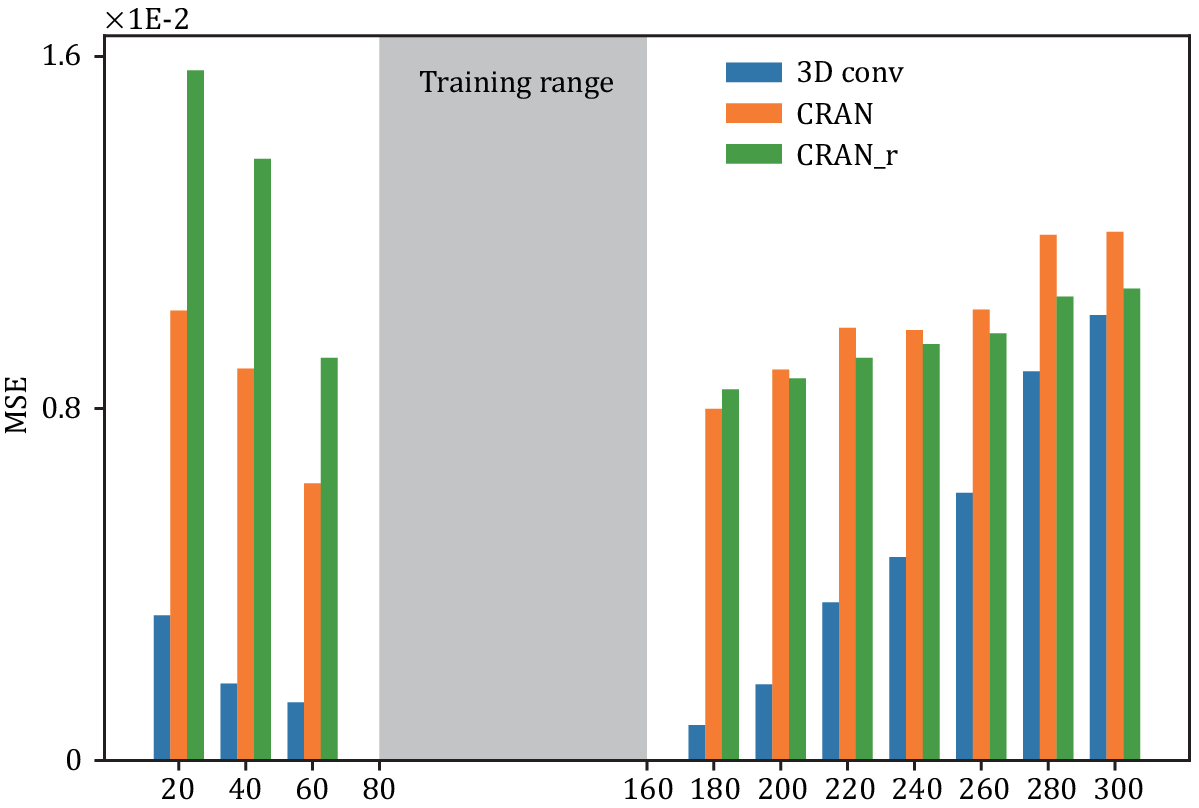}
    \caption{Temporal averaged MSE vs Reynolds number for CRAN, CRAN\_r and 3D convolution.}
\label{fig:new}

\end{figure*}

In Fig. \ref{fig:new}, we plot the temporal averaged MSE from the three models to analyze the results in the extrapolation regime and the impact of varying Reynolds number on the models. 
The temporal mean squared error for the 3D convolution model is five times smaller than the CRAN architectures close to the training range parameter, as shown in Fig. \ref{fig:new}.
In the extrapolation regimes close to the training range, the 3D convolution network outperforms the encoder-propagator-decoder architecture based on the CRAN model.

\subsection{Discussion}
Dynamical systems and unsteady partial differential equations have dependent variables that are a function of both time and space. Many numerical methods have been designed, which rely on the discretization of space and then solving the resulting ODEs. Traditional deep learning-based models are attempting to solve  PDEs following this strategy.  Deep learning algorithms employ an encoder network to capture the dominant spatial dimension and then project the high-dimensional coordinates on the dominant spatial modes.
The algorithms employ recurrent neural networks to evolve these low-dimensional trajectories and recover the high-dimensional physical space using the decoder network. This strategy of encoding-propagating-decoding for time-dependent partial differential equations leads to modeling error and reduces the extrapolation capability.
A combined space-time method based on a 3D convolution network solves the time-dependent partial differential equation over a space-time cylinder in a unified manner. The \emph{implicit bias} of learning coupled space-time correlation allows to mitigate the modeling error of the encoder-propagator-decoder network and hence allows it to perform better in the extrapolation regime.

\section{Conclusions}
\label{conclusion}
In this paper, we have presented a novel method for learning a combined space-time reduced-order model for predicting fluid flow. 
The model employs a 3D convolution architecture for the task of learning coupled spatio-temporal correlation. 
The model learns from the vortex dynamics data and predicts the physics of different regimes, i.e., the stable base flow with a low Reynolds number and the vortex shedding behind a circular cylinder.
We demonstrated that the coupled spatial-temporal correlations can serve as powerful implicit bias in the deep neural network hence they can provide better parameter space generalization.
Our model outperforms a current state-of-the-art deep learning-based reduced order method that employs an encoder-propagator-decoder architecture.
The proposed 3D convolution framework is general and has the potential to be used for predicting large-scale 3D fluid flow and fluid-structure interaction problems.

\section*{Acknowledgements}
The authors would like to acknowledge the funding support from the University of British
Columbia (UBC) and the Natural Sciences and Engineering Research Council of Canada (NSERC).
This research was supported in part through computational resources and services provided by Advanced Research Computing (ARC) at the University of British Columbia and Compute Canada.
\section*{Nomenclature}
\noindent\begin{longtable}{@{}l @{\quad=\quad} l@{}}
ADAM  & Adaptive Moment Estimation \\
CNN &    Convolutional Neural Network \\
CRAN & Convolutional Recurrent Autoencoder Network \\
FOM   & Full-Order Model \\
K & Convolution Kernel \\
$\mathcal{L}$ & Loss Function \\
LSTM & Long Short-term Memory \\
MSE & Mean Squared Error \\
N & Full-order Spatial Dimensions \\
$\mathcal{N}_{e}(x)$ & Neighbourhood of $x$ \\
$N_T$ & Total Temporal Length \\
$N_m$ & Number of Reynolds number sample\\
$N_t$ & Input Sequence Length \\
P & Pooling Operator \\
PDE & Partial Differential Equation \\
r & Reduced Dimension \\
$Re$ & Reynolds Number \\
RNN  & Recurrent Neural Network \\
ROM  &  Reduced-order Model \\
s & Stride Length \\
$\mathbf{U}_N$ & Full-order Solution \\
$\mathbf{U}_r$ & Reduced-order Solution \\
X & Model Input\\
Y & Ground Truth\\
$\sigma$ & Non-linear Activation \\
$\Psi_E$ & Encoder Network \\
$\Psi_D$ & Decoder Network \\
$\theta$ & Trainable Parameters\\
$< >$ & Temporal-average \\
$\overline{(.)}$ & Min-max Normalized Data \\
$(\hat{.})$ & Time-evolved Prediction \\

\end{longtable}

\section*{Data Availability}
The data that support the findings of this study are available from the corresponding author upon reasonable request.
\section*{Conflict of interest}
The authors declare that they have no conflict of interest.

\section*{References}
\bibliographystyle{plain}
\bibliography{mybibfile}

\end{document}